\documentclass{webofc}
\usepackage[varg]{txfonts}   
\usepackage{graphicx} 
\usepackage{url}

\begin{document}
\title{Repurposing of the Run 2 CMS High Level Trigger \protect\\ Infrastructure as a Cloud Resource for Offline Computing}

\author{
\firstname{Marco} \lastname{Mascheroni}\inst{1} \thanks{\email{marco.mascheroni@cern.ch}} \and
\firstname{Antonio} \lastname{Pérez-Calero Yzquierdo}\inst{2,3} \thanks{\email{aperez@pic.es}} \and
\firstname{Edita} \lastname{Kizinevic}\inst{4} \and
\firstname{Farrukh Aftab} \lastname{Khan}\inst{5} \and
\firstname{Hyunwoo} \lastname{Kim}\inst{5} \and
\firstname{Maria} \lastname{Acosta Flechas}\inst{5} \and
\firstname{Nikos} \lastname{Tsipinakis}\inst{4} \and
\firstname{Saqib} \lastname{Haleem}\inst{6} \and
\firstname{Damiele} \lastname{Spiga}\inst{7} \and
\firstname{Christoph} \lastname{Wissing}\inst{8} \and
\firstname{Frank} \lastname{Würthwein}\inst{1}~on behalf of the CMS Collaboration}

\institute{
University of California San Diego, La Jolla, CA, USA \and
Centro de Investigaciones Energ\'eticas, Medioambientales y Tecnol\'ogicas (CIEMAT), Madrid, Spain \and
Port d'Informaci\'o Cientifica (PIC), Barcelona, Spain \and
European Organization for Nuclear Research (CERN), Geneva, Switzerland \and
Fermi National Accelerator Laboratory, Batavia, IL, USA \and
National Centre for Physics, Islamabad, Pakistan \and
INFN Sezione di Perugia, Perugia, Italy \and
DESY, Hamburg, Germany
         }

\abstract{%
The former CMS Run 2 High Level Trigger (HLT) farm is one of the largest contributors to CMS compute resources, providing about 25k job slots for offline computing. This CPU farm was initially employed as an opportunistic resource, exploited during inter-fill periods, in the LHC Run 2. Since then, it has become a nearly transparent extension of the CMS capacity at CERN, being located on-site at the LHC interaction point 5 (P5), where the CMS detector is installed. This resource has been configured to support the execution of critical CMS tasks, such as prompt detector data reconstruction. It can therefore be used in combination with the dedicated Tier 0 capacity at CERN, in order to process and absorb peaks in the stream of data coming from the CMS detector. 
The initial configuration for this resource, based on statically configured VMs, provided the required level of functionality. However, regular operations of this cluster revealed certain limitations compared to the resource provisioning and use model employed in the case of WLCG sites. A new configuration, based on a vacuum-like model, has been implemented for this resource in order to solve the detected shortcomings. This paper reports about this redeployment work on the permanent cloud for an enhanced support to CMS offline computing, comparing the former and new models’ respective functionalities, along with the commissioning effort for the new setup.
}
\maketitle

\section{Introduction}
\label{sec:intro} 

The High-Level Trigger (HLT) system within the CMS experiment~\cite{cms} at CERN played a pivotal role in the LHC Run 2 by assisting in the selection of collision events, thereby advancing our comprehension of particle physics. As we transition from Run 2 to Run 3, the CMS experiment is adapting its resources for more efficient data processing.

\begin{figure}[ht]
\begin{center}
\includegraphics[width=10cm]{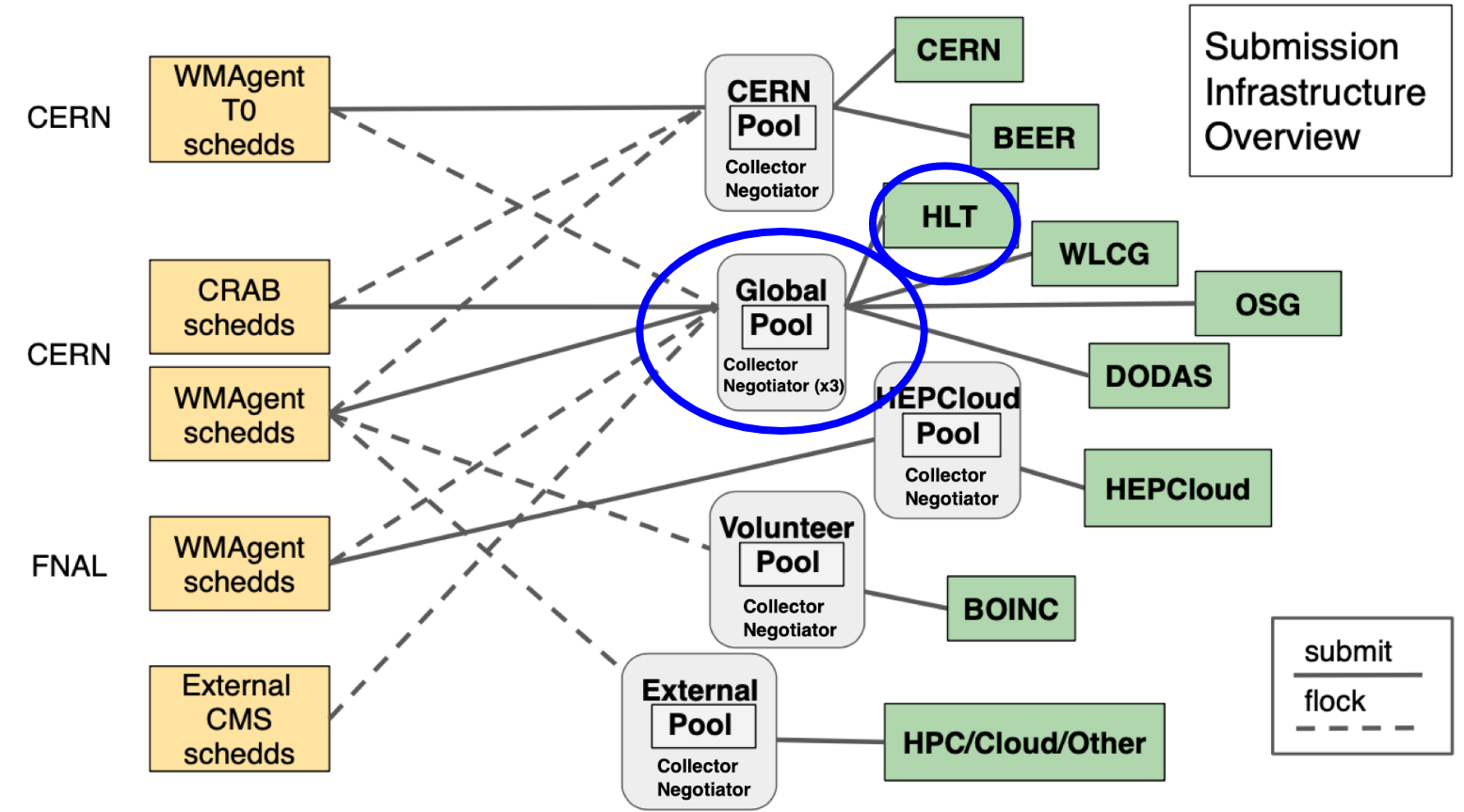}
\caption{CMS SI current configuration, including multiple federated pools of resources allocated from diverse origins (green boxes) and sets of distributed job schedulers (schedds), handling the Tier 0 and centralized production workloads (WMAgent), as well as analysis job submission (CRAB).} 
\label{fig:complexity}
\end{center}
\end{figure}

The CMS Online Cloud (deployed on the HLT farm resources) was commissioned during Run 2 to dedicate HLT resources for offline data processing when not needed to support online data taking~\cite{old_model}. Offline computing takes place in the Global pool~\cite{globalpool}, one of the federated HTCondor~\cite{htcondor} pools managed by the Submission Infrastructure team in CMS Computing. This set of federated pools are created using GlideinWMS~\cite{gwms} and collectively form a distributed and dynamic computing environment, capable of efficiently processing the ever-expanding volume of data generated by the CMS experiment at CERN, as shown in Figure \ref{fig:complexity}.

\begin{figure}[ht]
\begin{center}
\includegraphics[width=10cm]{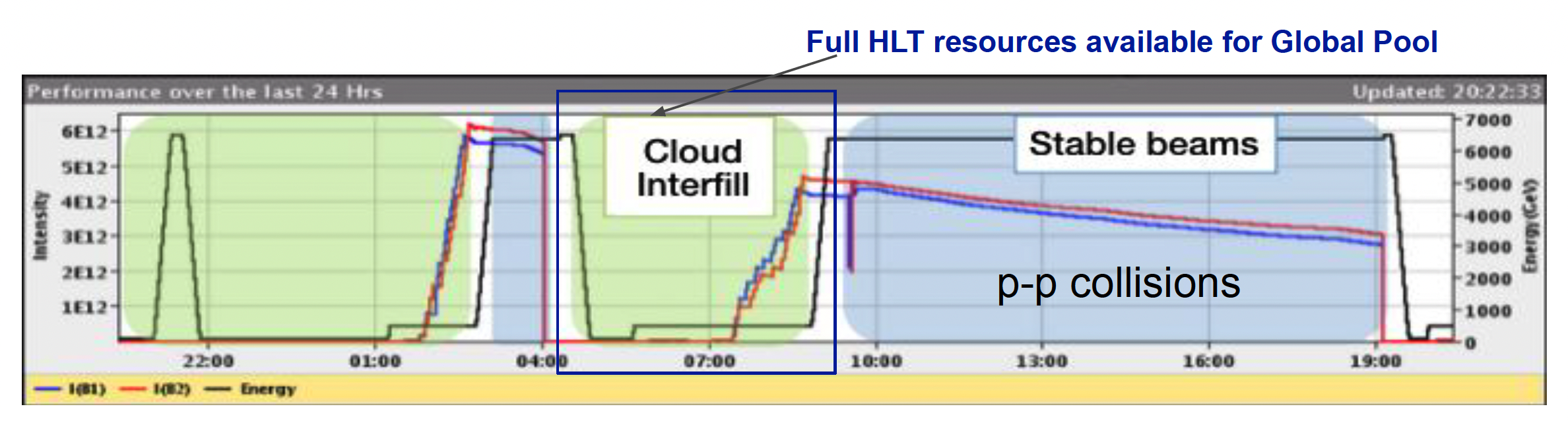}
\caption{LHC operation cycles and interfill operations of the HLT farm for CMS offline computing tasks.} 
\label{fig:interfill}
\end{center}
\end{figure}

OpenStack virtual machines (VMs) equipped with HTCondor Startd daemons are initiated to utilize available CPU resources during both Interfill and Fill modes, see Figure \ref{fig:interfill}. When the High-Level Trigger (HLT) requires resources, these VMs are gracefully suspended. The \emph{MaxHibernationTime} HTCondor knob is used to allow a delay of 24 hours at both the Central Control Board (CCB) and Workload Management Agents (WMAgents) schedds. This delay ensures that suspended jobs can be seamlessly resumed within the CMS Global pool when the VMs are reintegrated, a process triggered by diminished online processing demands


\begin{figure}[ht]
\begin{center}
\includegraphics[width=8cm]{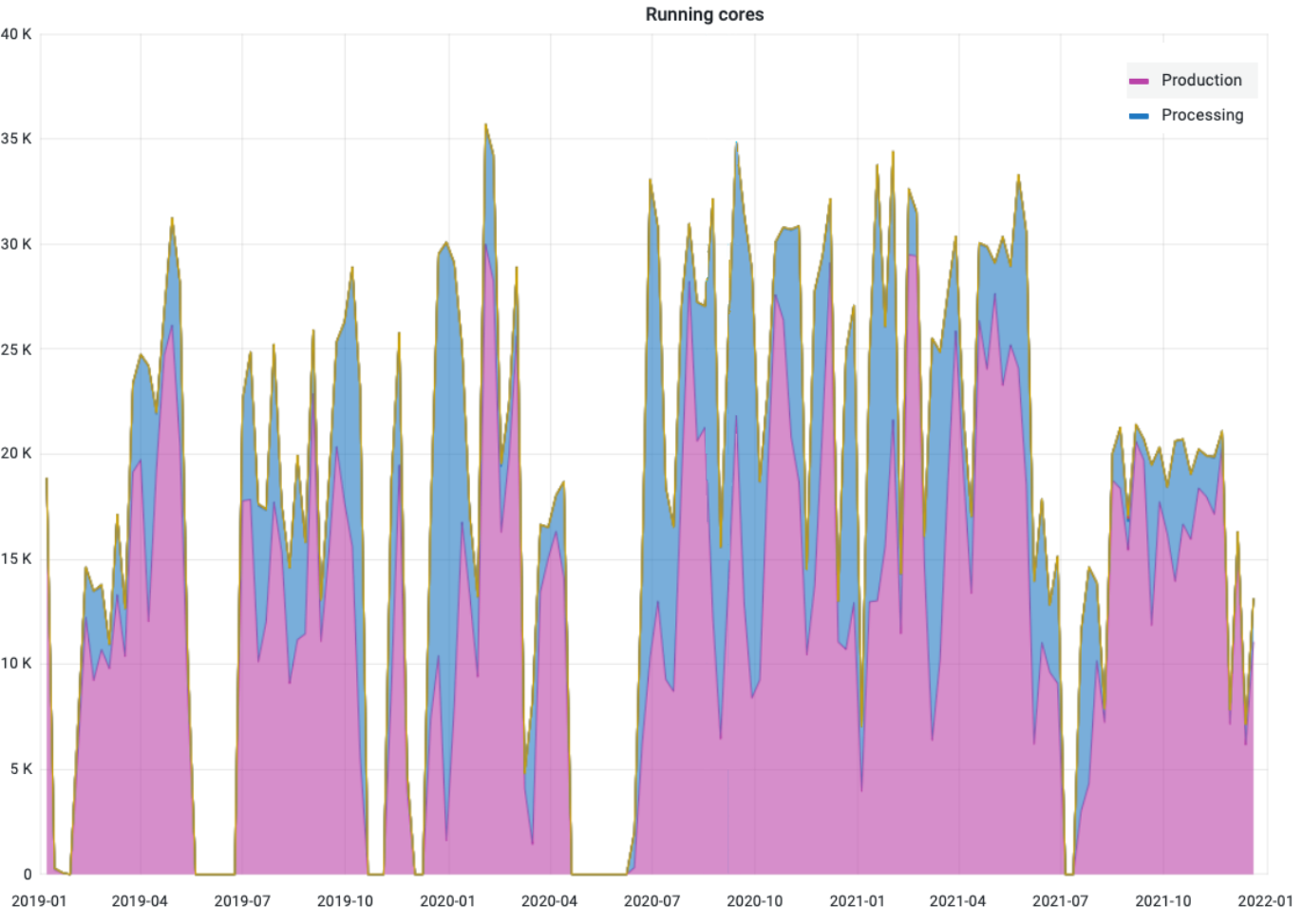}
\caption{Use of HLT farm resources in support of CMS offline computing tasks during the LS2 years.} 
\label{fig:LS2}
\end{center}
\end{figure}

Figure \ref{fig:LS2} shows the usage of the HLT farm during Run 2. The farm was intensively used as opportunistic resource for offline processing.

\section{The HLT farm Hardware status for Run 3}
\label{sec:hardware} 
As Run 3 approached, a new computing farm was commissioned for online usage. Currently, CMS benefits from the availability of two cloud-based High-Level Trigger (HLT) farms:
\begin{itemize}
    \item Run 2 HLT farm (aka Permanent cloud): This farm, boasting approximately 25,000 CPU cores, has been fully repurposed to support offline computing. It operates under the new CMS site name, T2\_CH\_CERN\_P5. While the hardware is out of warranty, it remains within the constraints of the current Data Acquisition (DAQ) and HLT power budget at the data center, with ample space available. As such, it is slated for use until the conclusion of Run 3, though server repairs are subject to best-effort and resources may diminish over time. The decommissioning of this farm is scheduled for the end of Run 3.
    
    \item This opportunistic cloud comprises approximately 33,000 CPU cores and will be utilized for offline computing during periods when the LHC is not actively taking data. The CMS site name T2\_CH\_CERN\_HLT has been reserved for this new HLT farm. Starting from Q1 2022, the new HLT nodes in this farm are equipped with dual NVidia T4 GPUs. These resources are expected to remain in use for trigger purposes until the conclusion of Run 3. Subsequently, they are projected to transition into a permanent cloud resource dedicated to offline computing.
\end{itemize}

\section{Original configuration of Run 2 HLT farm}
\label{sec: oldmodel} 

\begin{figure}[ht]
\begin{center}
\includegraphics[width=8cm]{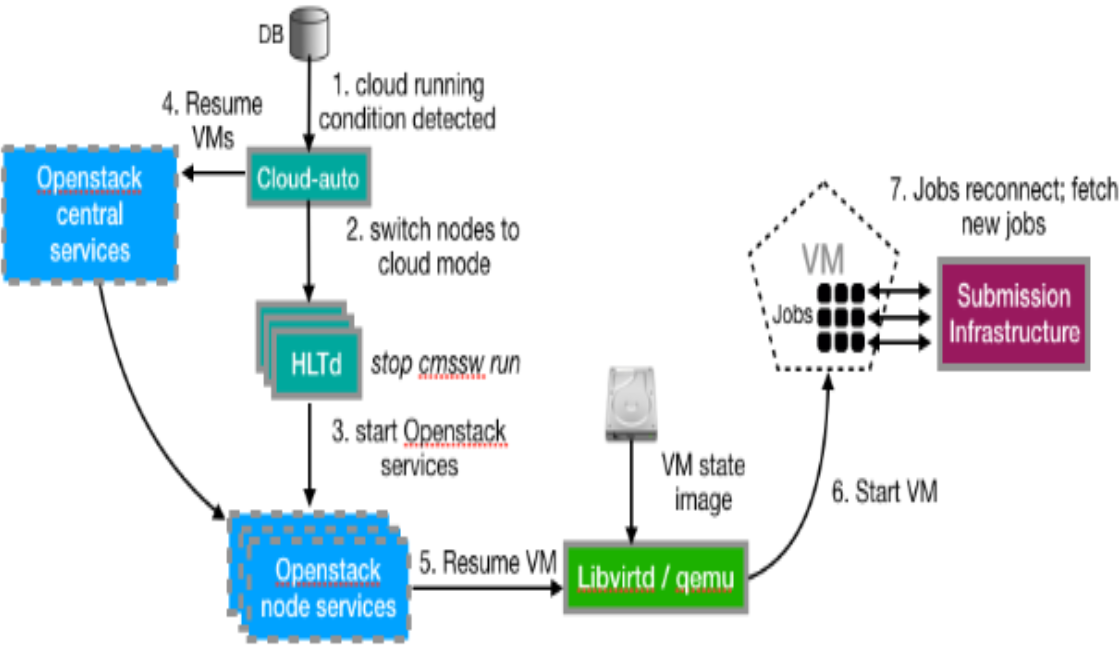}
\caption{Schematic view of the new HLT farm deployment model, based on vacuum-like pilots.} 
\label{fig:old_model}
\end{center}
\end{figure}

In the old configuration the HLT resources were dynamically repurposed for offline usage through the utilization of \emph{cloud auto} and \emph{HLTd} daemons, as illustrated in Figure \ref{fig:old_model}. Virtual machine (VM) images are crafted with a static HTCondor configuration provided by the Submission Infrastructure (SI) team. These configurations enable HTCondor Startd daemons to maintain persistent connections with the CMS pool, allowing them to operate for extended durations, often spanning weeks, in stark contrast to the typical 48-hour lifespan of regular glideins. Any adjustments, optimizations, or modifications to these configurations necessitate coordination with the SI team and the creation of updated VM images.

However, there are some notable issues and limitations associated with this approach. The prolonged existence of HTCondor Startd daemons raises concerns regarding configuration synchronization with the broader SI setup. Instances of outdated singularity wrapper scripts pointing to obsolete or missing EL8 images have been observed. Additionally, there is a lack of resource validation prior to these daemons connecting with the pool, resulting in a notable number of job failures attributable to cvmfs~\cite{cvmfs} issues, for example.

Furthermore, the long-lived HTCondor slots inadvertently contribute to resource allocation inefficiencies, particularly in the context of slot defragmentation. The absence of slot rotation with long-lived slots means that high-priority jobs, such as 8-core tasks like Tier 0 PromptReco, often remain in the queue for extended periods. To address this, an induced defragmentation mechanisms, facilitated by the HTCondor defrag daemon, was introduced and fine-tuned. Such measures are unnecessary for pilots with approximately 48-hour lifespans.

Lastly, it's important to note that the SI team lacks direct control over the tuning of HLT slot parameters in this approach. Any interventions or adjustments in this regard necessitate coordination with the DAQ team.

\section{Glideins in a Vacuum}
\label{sec: newmodel} 

To address the limitations of the old approach, the \emph{Glideins in a Vacuum} method was introduced. This approach enables the direct launching of Glideins from virtual machines (VMs) without the need for any intermediate compute elements or batch systems to manage pilot jobs onto worker nodes. This concept has garnered success within the WLCG context~\cite{vacuum_pilots}. The configuration of these Glideins is aligned aligns with the standard Global Pool Glideins, resulting in several integrated advantages. Notably, the relatively short-lived pilots provide a continuous and natural renewal of defragmented slots, effectively mitigating resource starvation issues (controlled by GLIDEIN\_Max\_Walltime). Furthermore, these Glideins undergo resource validation before connecting to the pool, contributing to a reduction in operational costs and increased control. Moreover, new configurations are automatically applied to pilots upon Frontend (FE) or factory reconfiguration.

To better reflect the dedicated nature of the old resources for offline computing, and to accommodate the introduction of Glideins in a vacuum, a new site name/endpoint, 'T2\_CH\_CERN\_P5,' was established. A new \emph{Glidein-launcher.service} script has been developed and has seamlessly been integrated into VMs through a contextualization script without necessitating modifications to the original VM image. The Glidein-launcher service, a systemd-based service, retrieves the wrapper script from the GWMS FE whenever it undergoes updates, subsequently initiating the launch of Glideins. The Submission Infrastructure (SI) team generates and publishes the wrapper for 'T2\_CH\_CERN\_P5' using a post-reconfigure hook of GlideinWMS Frontend.

\begin{figure}[ht]
\begin{center}
\includegraphics[width=8cm]{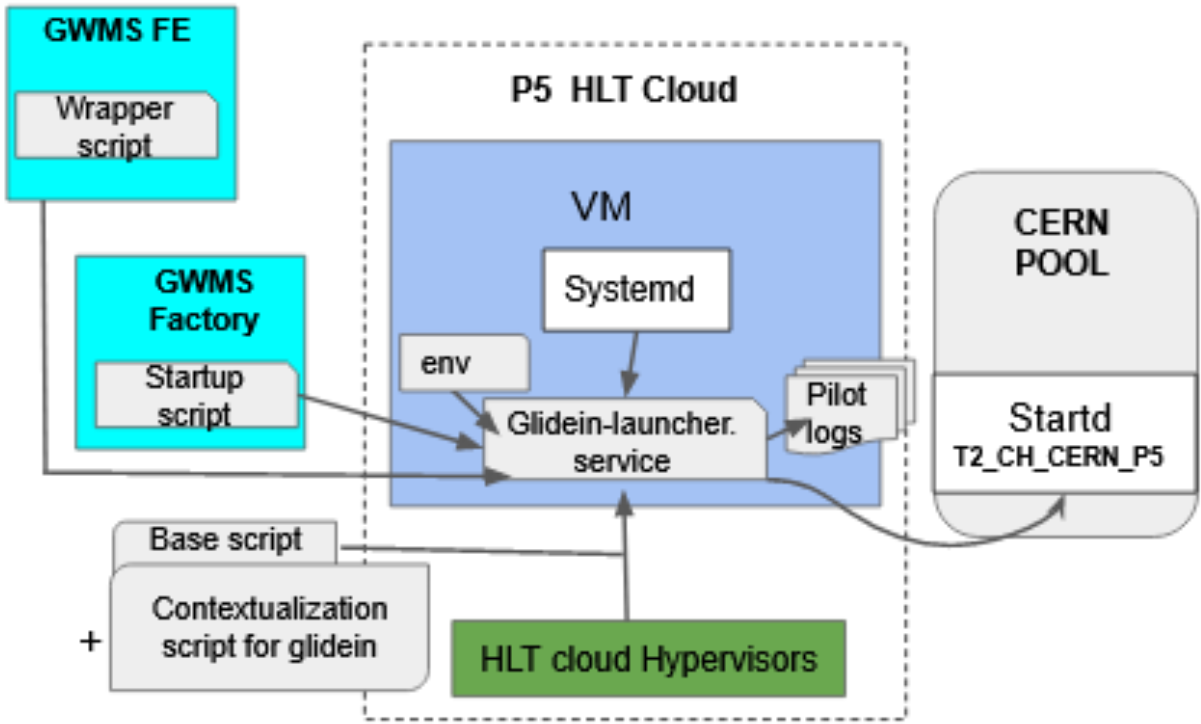}
\caption{Schematic view of the new HLT farm deployment model, based on vacuum-like pilots.} 
\label{fig:new_model}
\end{center}
\end{figure}

\begin{figure}[ht]
\begin{center}
\includegraphics[width=8cm]{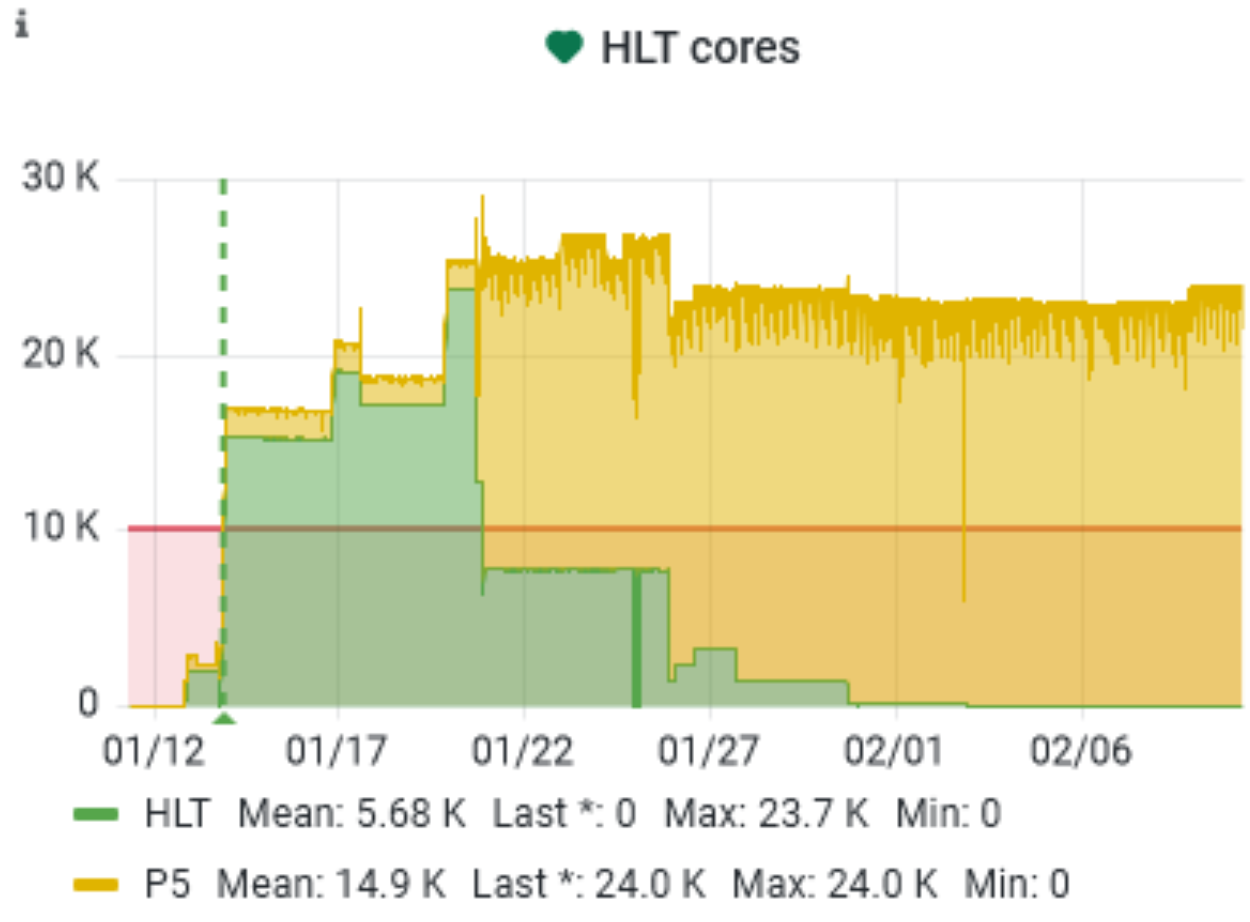}
\caption{Transition period for the deployment model of the HLT resources, based on vacuum-like pilots.} 
\label{fig:transition}
\end{center}
\end{figure}

The transition to the new Glideins-based site started in January 2023, with a complete switch to 'T2\_CH\_CERN\_P5' in February as shown in figure \ref{fig:transition}.

This transition empowers the SI team to swiftly configure, tweak, and optimize slots without the need for coordination with the DAQ team or the complexities of VM image updates. Examples of optimizations include parameters such as GLIDEIN\_Max\_Idle (set to 3600, or 1 hour), GLIDEIN\_Max\_Walltime (approximately 48 hours), GLIDEIN\_CPUS (configurable for whole node pilots), and MaxHibernateTime (set to 86400). Additionally, the system now allows only Production and Tier 0 jobs, a control facilitated through GlideinWMS Frontend. It introduces conditions like: 
\begin{verbatim}
ifthenelse(DESIRED_Sites is not undefined,
stringListMember(GLIDEIN_CMSSite, DESIRED_Sites), undefined)
\end{verbatim}
and
\begin{verbatim}
(GLIDEIN_CMSSite =!= 'T2_CH_CERN_P5' ||
WMAgent_AgentName =!= UNDEFINED)    
\end{verbatim}
to further refine job assignments and validations, thereby reducing the likelihood of job failures. Lower idle times for high-priority tasks has also been observed.

With the migration of 'T2\_CH\_CERN\_P5' from the Global to the CERN Pool, several advantages have emerged. This transition brings about a level of similarity with 'T2\_CH\_CERN' in terms of resource shares and priorities, encompassing Tier 0, production, and analysis jobs (excluding analysis jobs in P5). Furthermore, it effectively isolates 'T2\_CH\_CERN\_P5' from potential issues that may impact the global pool, enhancing resilience and resource stability for tasks primarily associated with Tier 0 operations. This migration also alleviates the pressure on the global pool collector and the global pool T2 negotiator, contributing to the overall robustness and efficiency of the system.

\section{Commissioning Tier 0 jobs in the reconfigured HLT farm in Run 3}
\label{sec:commissioning}

\begin{figure}[ht]
\begin{center}
\includegraphics[width=10cm]{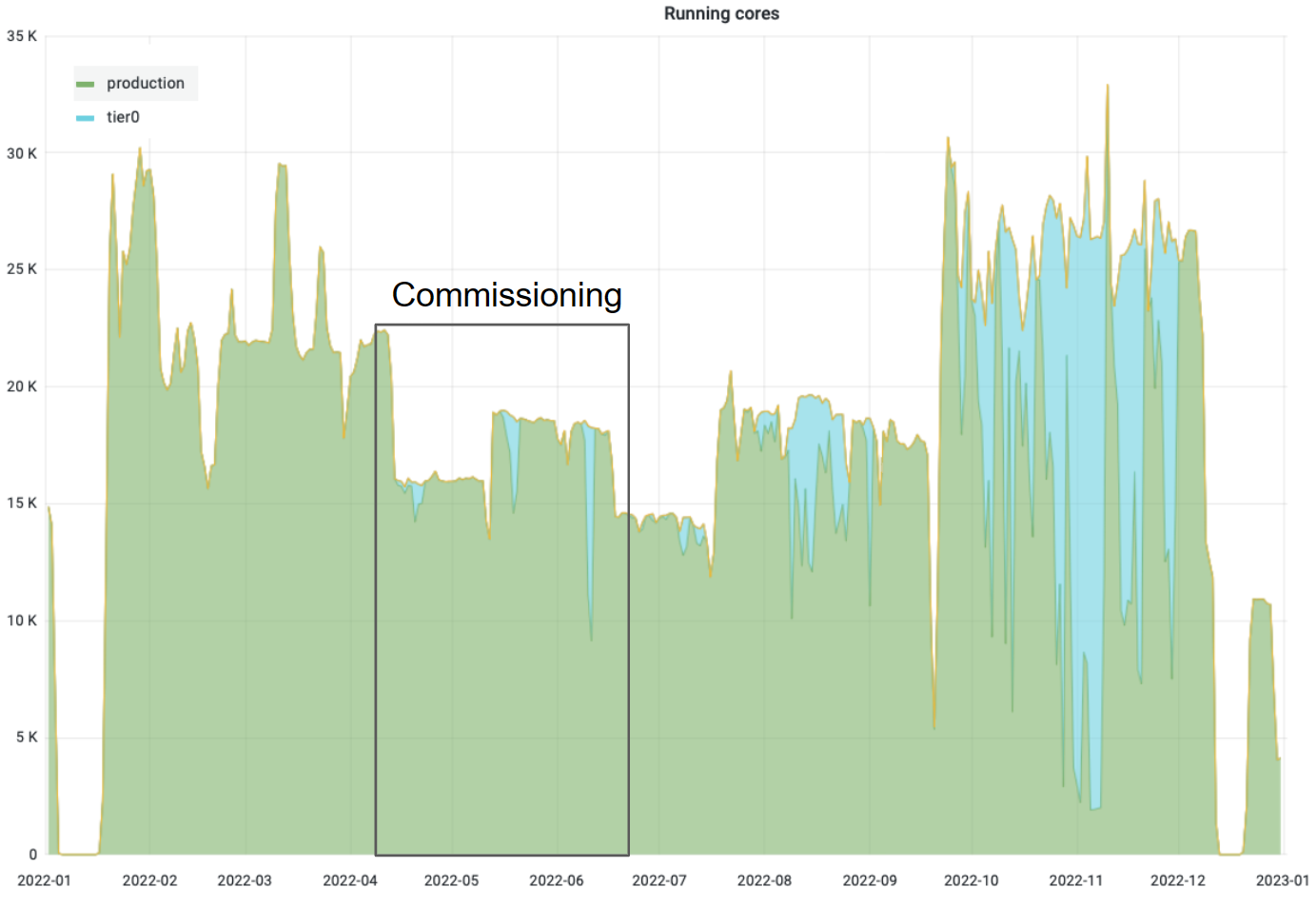}
\caption{Transition period for the deployment model of the HLT resources, based on vacuum-like pilots.} 
\label{fig:HLTForT0}
\end{center}
\end{figure}

In the initial year of LHC Run 3 (2022), the Run 2 HLT farm found new opportunities for utilization. This time, it extended its support to Tier 0 operations, specifically in the realm of prompt reconstruction tasks. To ensure a seamless transition, a commissioning phase was initiated in the spring of 2022, aimed at enabling Tier 0 tasks to efficiently access the established HLT farm and secure slots at a rate conducive to their needs, see Figure \ref{fig:HLTForT0}.

During this period, a notable challenge emerged: partitionable slots within long-lived virtual machines (VMs) often experienced fragmentation due to low-core-count production jobs. To address this issue and facilitate the execution of critical PromptReco tasks, which require 8-core slots, an induced defragmentation mechanism for the HLT slots was configured and optimized.

As soon as stable beams were declared, the resource seamlessly extended the Tier 0 capacity, responding to the demands imposed by the data acquisition rates. This adaptive use of the Run 2 HLT farm played a crucial role in supporting the continuous flow of data during LHC Run 3.

\begin{figure}[ht]
\begin{center}
\includegraphics[width=10cm]{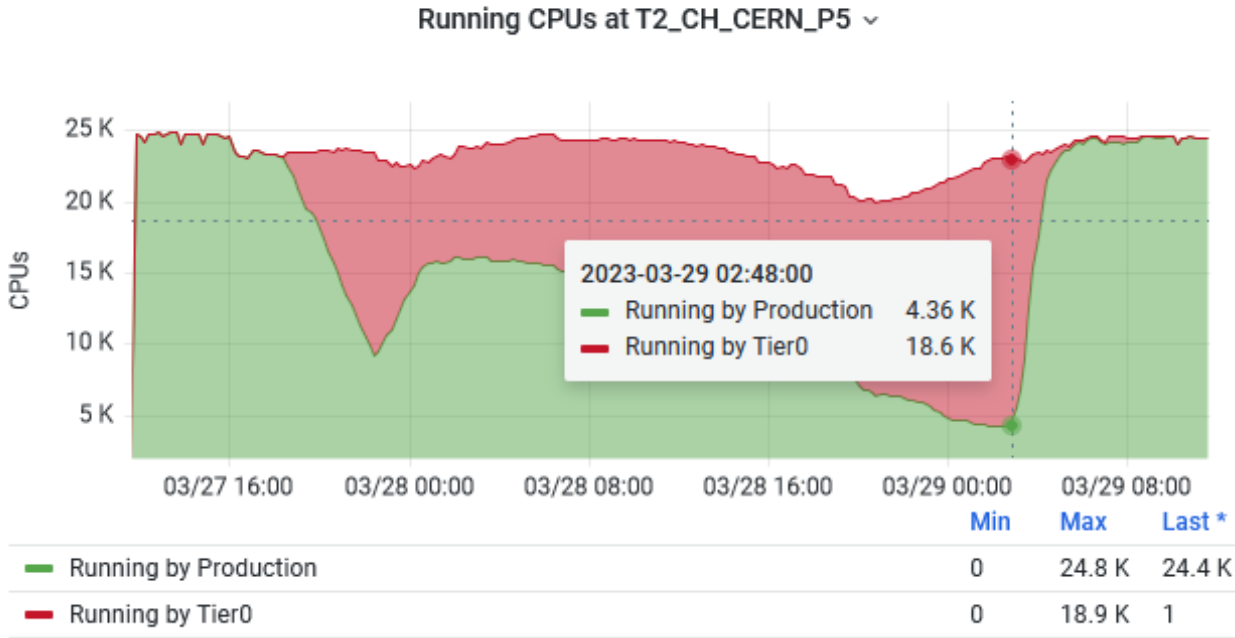}
\caption{Transition period for the deployment model of the HLT resources, based on vacuum-like pilots.} 
\label{fig:HLTT0Recent}
\end{center}
\end{figure}

In 2023, Tier 0 jobs are efficiently processed on T2\_CH\_CERN\_P5 following the successful transition to the new vacuum model. In 2022, the inclusion of the Run 2 HLT in the resource pool significantly bolstered the capacity for executing Tier 0 prompt reconstruction tasks. The commissioning of the new site name and glidein mechanism, specifically tailored to accommodate Tier 0 jobs, has undergone rigorous testing and validation in anticipation of the 2023 LHC data-taking season. With this new approach it was shown that the allocation of CPU cores for Tier 0, characterized by high-priority jobs, was really fast as shown in Figure \ref{fig:HLTT0Recent}.

\section{Conclusions and future work}
\label{sec:conclusions}

This paper has detailed the successful repurposing of the Run 2 CMS HLT infrastructure into a dedicated farm designed specifically for offline computing tasks. This transformation became feasible due to the deployment of a new HLT farm tailored for Run 3. We have explored the implementation of the Vacuum Model Glideins, an approach that provides enhanced flexibility for offline computing.

Our transition to a Glidein-based acquisition and management approach for P5 resources has yielded several identified advantages, aligning with our initial motivations for this shift. Since its full implementation in February, this approach has performed as anticipated, and we have already realized benefits such as automated pilot reconfiguration. We are now planning to extend this pilot-based methodology to the new Run 3 HLT farm, configuring Glideins with the necessary parameters to accommodate the intermittent nature of these resources. We will continue to utilize the reserved sitename 'T2\_CH\_CERN\_HLT' for this purpose. Presently, the opportunistic use of the new farm does not encompass GPU nodes due to pending configurations for GPU virtualization.

Additionally, we have examined the potential of leveraging the new HLT infrastructure in both Interfill and Fill modes to meet diverse offline usage requirements in the future. The Run 3 HLT farm will be opportunistically utilized during specific periods, notably during LHC technical stops and in what we refer to as the 'Interfill mode'. This mode is active during brief intervals, typically spanning only a few hours, when the LHC temporarily ceases the generation of physics data. The prerequisites for enabling Interfill mode for offline computing are already in place, including the capabilities for VM suspension and disk state preservation on the DAQ side. Similarly, the SI team has implemented a job suspension and resumption feature to facilitate Interfill mode utilization.

As part of ongoing efforts, a subset of nodes from the Run 3 HLT farm is designated for testing and seamless integration into the broader CMS offline resource pool. This approach ensures the smooth incorporation of these resources into our computational infrastructure, further enhancing our capacity for efficient offline data processing.

\section*{Acknowledgements}
This work was partially supported by the Spanish Ministry of Science and Innovation under grants PID2019-110942RB-C21, PID2019-110942RB-C22 and PID2020-113807RA-I00, which include FEDER funds from the European Union, and by the US National Science Foundation under Grant No. 2121686.

\end{document}